\documentclass[twocolumn,iop, aps,superscriptaddress,showpacs]{revtex4-1}

\usepackage{mathptmx}
\usepackage{subfigure}
\usepackage{dcolumn}
\usepackage{amsmath,amssymb}
\usepackage{bm}
\usepackage{color}
\usepackage{latexsym}
\usepackage{epstopdf}
\usepackage{color}
\usepackage[english]{babel}
\usepackage{latexsym}

\usepackage{psfrag,graphicx} 
\usepackage{epsf} 
\usepackage{subfigure} 
\usepackage{amsmath} 
\usepackage{amssymb} 
\usepackage{amsfonts}
\usepackage{bm}
\usepackage{natbib}
\usepackage{epstopdf}
\usepackage{appendix}
\usepackage{changes}

\definecolor{mygrey}{gray}{0.35}
\definecolor{myblue}{rgb}{0.2,0.2,0.8}
\definecolor{myzard}{cmyk}{0,0,0.05,0}
\definecolor{mywhite}{rgb}{1,1,1}
\definecolor{myred}{rgb}{1,0.,0.3}

\usepackage[colorlinks=true,citecolor=myblue,linkcolor=myred]{hyperref}

\def\be{\begin{equation}}
\def\ee{\end{equation}}
\def\ba{\begin{align}}
\def\enda{\end{align}}
\def\bi{\begin{itemize}}
\def\ei{\end{itemize}}

 \def\ee{\mathord{\rm e}}

 \def\ee{\mathord{\rm e}}

\renewcommand{\ee}{{\rm e}}

\def\beq{\begin{equation}}
\def\beq{\begin{equation}}
\def\eeq{\end{equation}}

\begin{document}

\title[Short Title]{Control methods for improved Fisher information with quantum sensing}

\author{Tuvia Gefen}
\affiliation{Racah Institute of Physics, The Hebrew University of Jerusalem, Jerusalem 
91904, Givat Ram, Israel}
\author{Fedor Jelezko}
\affiliation{Institute for Quantum Optics, Ulm University, Albert-Einstein-Allee 11, Ulm 89081, Germany}
\author{Alex Retzker}
\affiliation{Racah Institute of Physics, The Hebrew University of Jerusalem, Jerusalem 
91904, Givat Ram, Israel}
\date{\today}


\begin{abstract}
{Recently new approaches for sensing the frequency of time dependent Hamiltonians have been presented, and it was shown that the optimal Fisher information scales as $T^{4}.$ We present here our interpretation of this new scaling, where the relative phase is accumulated quadratically with time, and show that this can be produced by a variety of simple pulse sequences. Interestingly, this scaling has a limited duration, and we show that certain pulse sequences prolong the effect. The performance of these schemes is analyzed and we examine their relevance to state-of-the-art experiments. We analyze the $T^{3}$ scaling of the Fisher information which appears when multiple synchronized measurements are performed, and is the optimal scaling in the case of a finite coherence time.    }
\end{abstract}
\maketitle

\section{Introduction}
\subsection{Background}
 Quantum sensing\cite{degen2016,budker2007optical,rondin2014magnetometry} and metrology\cite{bollinger1996optimal,giovannetti2004quantum,giovannetti2006quantum} 
utilizes individual quantum systems or collective quantum phenomena to improve measurement precision.
A typical problem in these fields is formulated as follows: A Hamiltonian $H$ depends on a parameter $g,$ that we wish to  estimate. In order to gain information about $g,$ we initialize the quantum state of the probe, let it evolve for a duration of $\tau$ and perform a measurement in the end. The outcome then depends on $g$ and by iterating this process several times we are able to estimate $g.$

The precision is then determined according to the standard deviation of the estimator, which is bounded by $\frac{1}{\sqrt{I_{g}}},$ where $I_{g}$ is the Fisher information (FI) \cite{cramer}. 
Since Bayesian estimation saturates this bound, the precision is quantified by the FI. Given the probabilities of the different outcomes $\left\{ P_{i}\right\} _{i},$ the FI reads: $I_{g}=\underset{i}{\sum}\frac{\left(\frac{dP_{i}}{dg}\right)^{2}}{P_{i}}.$ 
In the context of quantum measurements, given the final state of the probe, the FI is determined by the measurement performed at the end. The maximal FI, achieved with an optimal measurement, is termed as QFI and is given (for a pure state $|\psi_{g} \rangle$) by \cite{wootters,caves}: 
\begin{equation}
4\left(\langle\partial_{g}\overset{}{\psi}_{g}|\partial_{g}\overset{}{\psi}_{g}\rangle-|\langle\overset{}{\partial_{g}\psi}_{g}|\psi_{g}\rangle|^{2}\right)=4\text{Var}\left(iU^{\dagger}\partial_{g}{U}\right),
\label{QFI}
\end{equation}
where $U$ is the time evolution unitary and the derivative is with respect to $g.$
A standard sensing scenario is measurement of the strength of time independent signal, i.e.  $H=g \sigma_{z},$ where $g$ is to be estimated. A typical Ramsey sensing scheme would then include initialization of the probe to an eigenstate of $\sigma_{X}$ (or any other operator in $\sigma_{X}$-$\sigma_{Y}$ plane). 
During the free evolution  a relative phase of, $\phi=2 g t,$ is accumulated between the different eigenstates of $\sigma_{Z},$ or equivalently a rotation by an angle of $\phi$ around $Z$ axis in Bloch sphere. The resulting QFI then reads: $\left(\frac{d\phi}{dg}\right)^{2}=4t^{2},$ and therefore $\Delta g=\frac{1}{2t}.$ 
The quadratic dependence of the FI on time is thus due to the linear growth of the relative phase with time.     
A finite coherence time, due to different kinds of noise, would usually lead to a precision that scales as $\frac{1}{\sqrt{TT_{2}}}$\cite{itano1993quantum,helstrom1969quantum,holevo2011probabilistic} ,where $T_{2}$ denotes the coherence time of the probe and $T$ is the total time.\\       
\subsection{Sensing the frequency of time dependent signals}
 A different sensing scenario, which is of great interest to nanoscale NMR and frequency standards, is the spectral reconstruction of a time dependent Hamiltonian. The simplest examples are given by $H_{1}=\Omega\left(\sigma_{X}\cos\left(2\omega t\right)+\sigma_{Y}\sin\left(2\omega t\right)\right)$ and $H_{2}=\Omega \sigma_{Z} \sin \left(    2\omega t      \right),$ where $\omega$ is to be estimated. Interestingly recent theoretical and experimental works have shown an improved behavior of the precision in that case \cite{Qdyne,pang2016quantum,pang2}. These studies reported that $T^{4}$ scaling of the FI of $\omega$ is achievable and a tight upper bound was calculated in ref. \cite{pang2016quantum}.
In this paper we present our interpretation of this scaling and a set of coherent control methods that lead to it. The paper is structured as follows: We present our derivation of known results, then we show different control methods that appear to be more efficient. Afterwards we briefly discuss the implications of an unknown initial phase of the signal, and the last section deals with the behavior of the FI when coherence time is limited.   


$T^{4}$ scaling of the FI of the frequency was first introduced in \cite{pang2016quantum} and was analyzed for the following Hamiltonian:
\begin{equation}
H_{1}=\Omega \left(  \sigma_{X}\cos\left(2 \omega t\right)+\sigma_{Y}\sin\left(2 \omega t\right) \right),
\label{H1}
\end{equation}
where $\omega$ is the parameter to be estimated.
The method proposed in \cite{pang2016quantum} for $T^{4}$ scaling can be understood as follows. Applying stroboscopically a control of:
\begin{equation}
  H_{1C}=-\Omega \left( \sigma_{X}\cos\left(2 \omega' t\right)+\sigma_{Y}\sin\left(2 \omega' t\right) \right)+\omega' \sigma_{Z},
  \end{equation}
 we get the following Hamiltonian:          
  \begin{eqnarray}
  \begin{split}
& H_{1e}=\Omega\sigma_{X}\left(\cos\left(2\omega t\right)-\cos\left(2\omega't\right)\right)+\\
&\Omega \sigma_{Y}\left(\sin\left(2\omega t\right)-\sin\left(2\omega't\right)\right)+\omega'\sigma_{Z}.
  \end{split} 
  \end{eqnarray}
  Now if $\omega$ is known to a high degree, namely $|\delta| t \ll 1,$ where $\delta=\omega-\omega',$ then after moving to the interaction picture with respect to $\omega' \sigma_{Z}$ we get:
  \begin{equation}
H_{1e} \approx 2\Omega\delta t\sigma_{Y}.    
  \end{equation}
  Now a standard Ramsey experiment would lead to a relative phase of:
  \begin{equation}
  \phi=2\int2\Omega\delta t\:dt=2\delta\Omega t^{2},
  \end{equation} 
  and the resulting QFI of $\delta$ reads: $\left(\frac{d\phi}{d\delta}\right)^{2}=4\Omega^{2}t^{4}.$ 
  This method, however, requires a knowledge of $\Omega.$

The Hamiltonian:
\begin{equation} 
H_{2}=\Omega \sigma_{Z} \sin \left( 2   \omega t      \right)
\label{Qdyne_Hamiltonian}
\end{equation}
was studied in ref. \cite{Qdyne}, as it was realized experimentally with NV centers in diamond. We can readily observe that a similar method also works for this case. 
Choosing a control Hamiltonian of:
\begin{equation}   
H_{2C}=-\Omega\sigma_{Z}\sin\left(2\omega't\right)+\frac{\pi}{2}\underset{N}{\sum}\delta\left(t-\frac{\pi}{4\omega'}\left(2N+1\right)\right)\sigma_{X},
\end{equation}
where $\delta\left(t-\frac{\pi}{4\omega'}\left(2N+1\right)\right)$ is the Dirac delta function, and the summation is over $N,$ namely the integer numbers. 
The term, $\frac{\pi}{2}\underset{N}{\sum}\delta\left(t-\frac{\pi}{4\omega'}\left(2N+1\right)\right)\sigma_{X},$ represents a decoupling sequence which is composed of fast $\pi$ pulses and can be realized by an XY8 sequence or a CPMG\cite{slichter2013principles,freeman1998spin}.
By adding this control to the original Hamiltonian we get:
\begin{equation}
H_{2e} \approx 2\Omega\delta t\sigma_{Z}\cos\left(2\omega't\right)+\frac{\pi}{2} \underset{N}{\sum} \delta\left(t-\frac{\pi}{4\omega'}\left(2N+1\right)\right)\sigma_{X}.
\end{equation}
 Now moving to the interaction picture with respect to the pulses we get: $H_{2e}=2\Omega\delta t\sigma_{Z}|\cos\left(2\omega't\right)|,$ which leads to a quadratic phase accumulation. 
However, as it was shown in \cite{Qdyne,pang2}, the first term in $H_{2C}$ is redundant and only pulses are required. By applying the following control Hamiltonian:
\begin{equation}
H_{2C}=\frac{\pi}{2}\underset{N}{\sum}\delta\left(t-\frac{\pi}{4\omega'}\left(2N+1\right)\right)\sigma_{X},
\end{equation}
namely just pulses, we obtain the following effective Hamiltonian:  
\begin{equation}        
H_{2e} \approx \frac{2}{\pi} \Omega \sin \left( 2 \delta t    \right) \sigma_{Z},   
\end{equation}  
where $\delta=\omega-\omega'.$ Assuming again that $|\delta| t \ll 1$ we get:
\begin{equation}  
H_{2e} \approx \frac{4}{\pi} \Omega \delta t \sigma_{Z}.   
\end{equation}  
Therefore a FI of $\left(\frac{4}{\pi}\right)^{2}\Omega^{2}t^{4}$ is achieved, it can be verified with the technique presented in ref. \cite{pang2016quantum} that this FI is optimal. Unlike the first method this method does not require knowledge of $\Omega.$ This implies that similar pulse sequences may lead to a phase acceleration with $H_{1}.$
Both of these methods, however, do require knowledge of the initial phase.
In the rest of the paper we refer to the Hamiltonian in eq. \ref{H1} as $H_{1}$ and to the Hamiltonian in  eq. \ref{Qdyne_Hamiltonian} as $H_{2}$.

\section {More efficient control methods for $H_{1}$} 
\subsection {First control method}
\label{first} 
\begin{figure}[!]
\begin{center}
\subfigure[]{\includegraphics[width=0.4\textwidth]{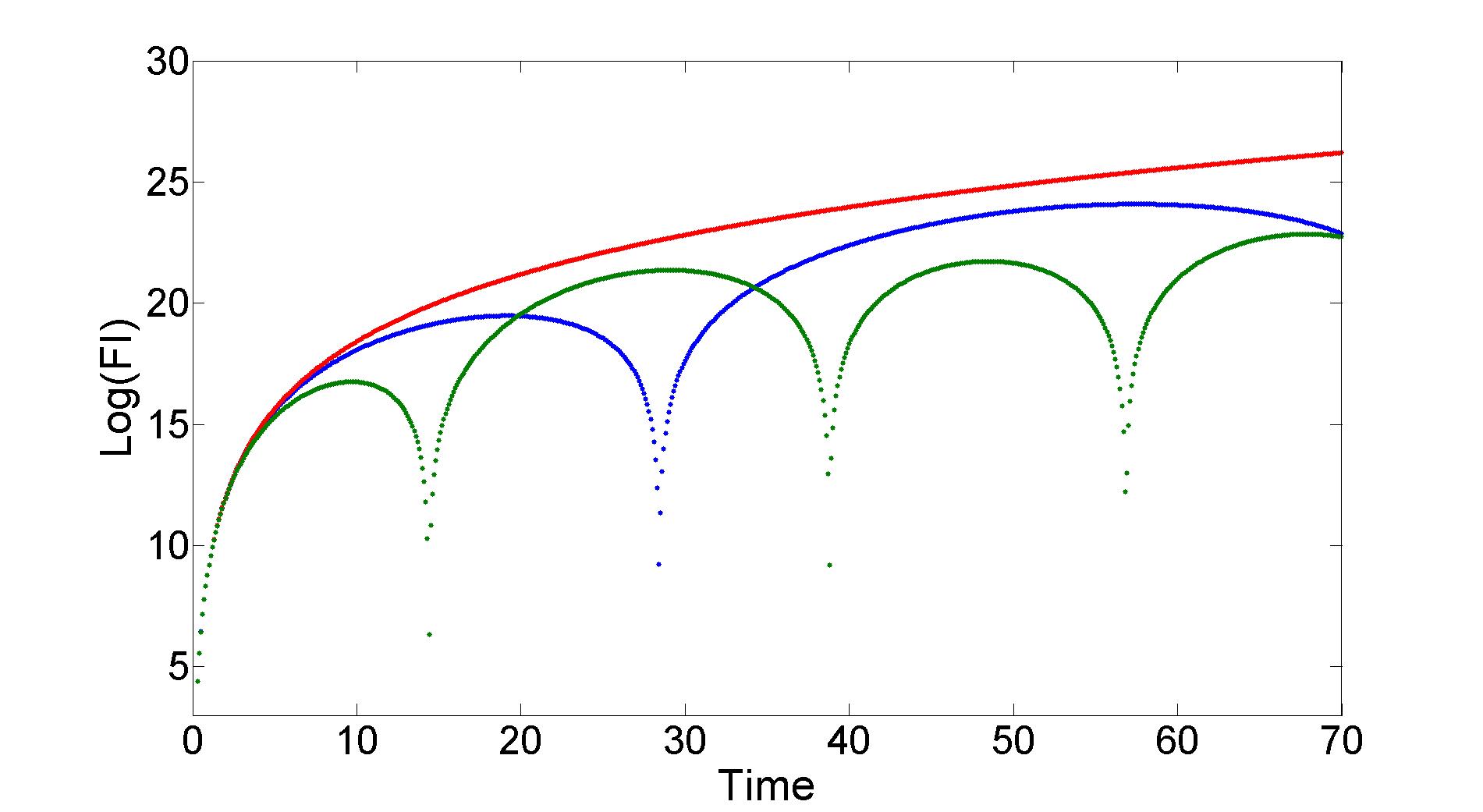}}
\subfigure[]{\includegraphics[width=0.4\textwidth]{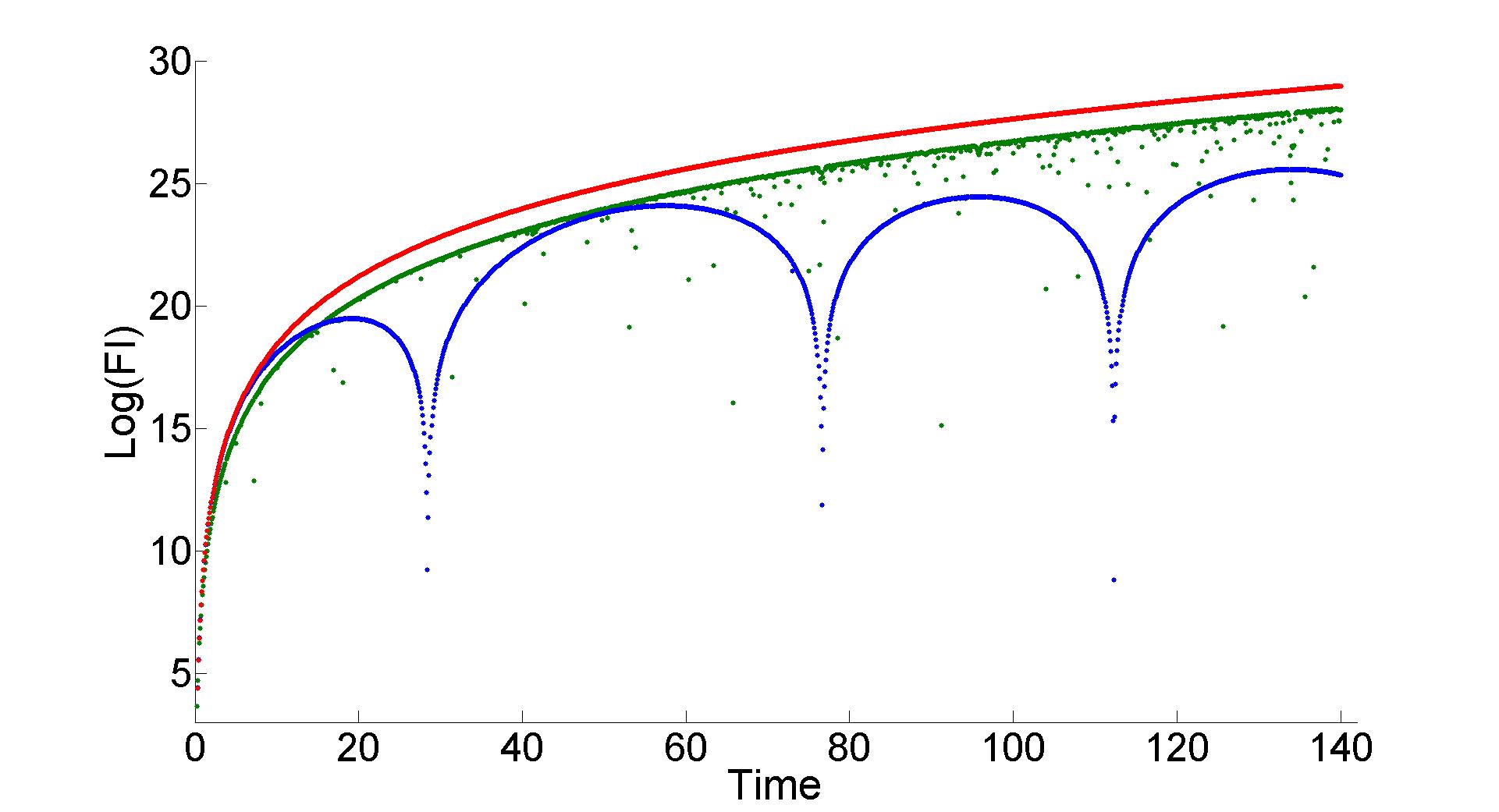}}
\end{center}
\caption{Top: FI achieved with the first control method (see section \ref{first}), compared with the optimal FI. The green curve corresponds to $\delta=0.08 \omega$ , the blue to $\delta=0.04 \omega$ and the red is the optimal FI. The optimal FI coincides with this method only for $\delta t \ll 1,$ as can be seen from eq. \ref{FI_method_1}. Bottom: Comparison between the two different control methods. The blue curve corresponds to the first method, the green to the second (see section \ref{second}) and the red is the optimal FI. For short times ($\delta t \ll 1$) the first method is superior as it achieves the optimal FI. However it loses this advantage quite quickly as it suffers from a shorter lifetime. In both plots $\Omega=50 \omega.$}
\label{comparison_methods}
\end{figure}

Concentrating on $H_{1}$,
we claim that phase acceleration can be obtained by a control method, which does not require knowledge of $\Omega$. Adding to $H_{1}$ a term of $H_{0}=\omega' \sigma_{Z},$ where $\omega'$ is our estimation of $\omega,$ and moving to the interaction picture with respect to it yields:
\begin{equation}
H_{1I}=\Omega\left(\sigma_{X}\cos\left(2\delta t\right)+\sigma_{Y}\sin\left(2\delta t\right)\right).
\label{interaction_Hamiltonian}
\end{equation}
Applying $\pi$-pulses of $\sigma_{Y}$ (in the interaction picture) every $\Delta t,$ such that $\Omega \Delta t, \delta \Delta t \ll 1,$ will transform $\sigma_{X}$ to $-\sigma_{X}$ so the term of $\Omega \sigma_{X} \cos \left( 2 \delta t \right)$ will be canceled out and we will be left with: $H_{1e}=\Omega \sigma_{Y}\sin \left(2 \delta t   \right).$ 
Hence a quadratic phase accumulation is achieved in the limit of $\delta t \ll 1.$ We remark that this dynamical decoupling could be implemented continuously by opening a large energy gap in $\sigma_{Y}$ direction.  
As reported in \cite{Qdyne} the FI of $H_{Ie}$ reads:
\begin{equation}
\frac{\Omega^{2}}{\delta^{4}}\left(\cos\left(2\delta t\right)-1+2\delta t\sin\left(2\delta t\right)\right)^{2},
\label{FI_method_1}
\end{equation}
therefore in the limit of $\delta t \ll 1$ the optimal FI is achieved. Eq. \ref{FI_method_1} explicitly shows that the lifetime of the $T^{4}$ scaling goes as $\sim \frac{1}{\delta} ,$ and for longer times the FI becomes $4\frac{\Omega^{2}}{\delta^{2}}\sin^{2}\left(2\delta T\right)T^{2}.$ This is illustrated with numerical results in fig. \ref{comparison_methods}(a). It means that favorable scaling is achieved only when the frequency of the signal is known with a very good precision and as the scheme is adaptive in nature this restricts the time of each measurement and lowers sensitivity.

It is natural to inquire whether other dynamical decoupling methods can achieve this scaling with a longer duration.

\begin{figure*}
\begin{center}
\subfigure[]{\includegraphics[width=12 cm]{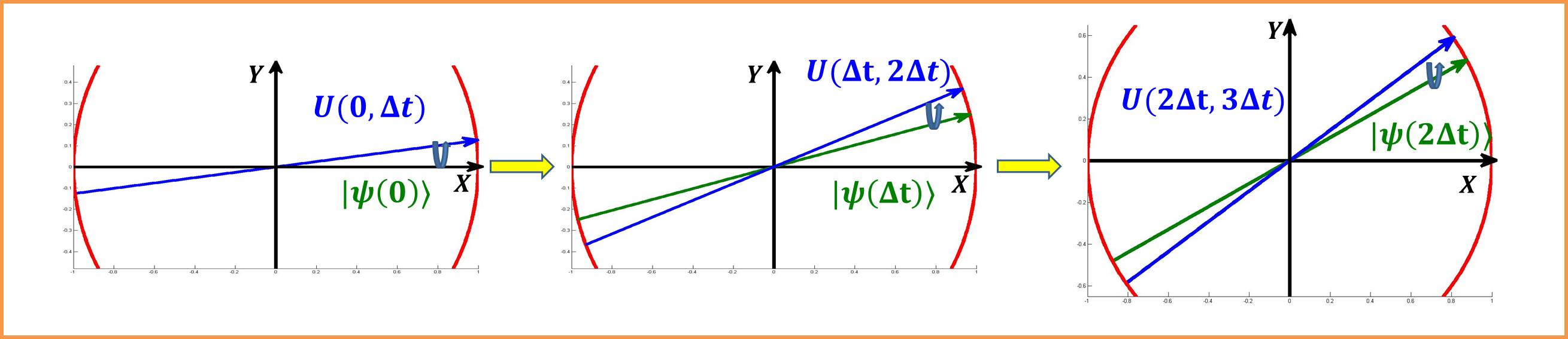}}
\subfigure[]{\includegraphics[width=13 cm]{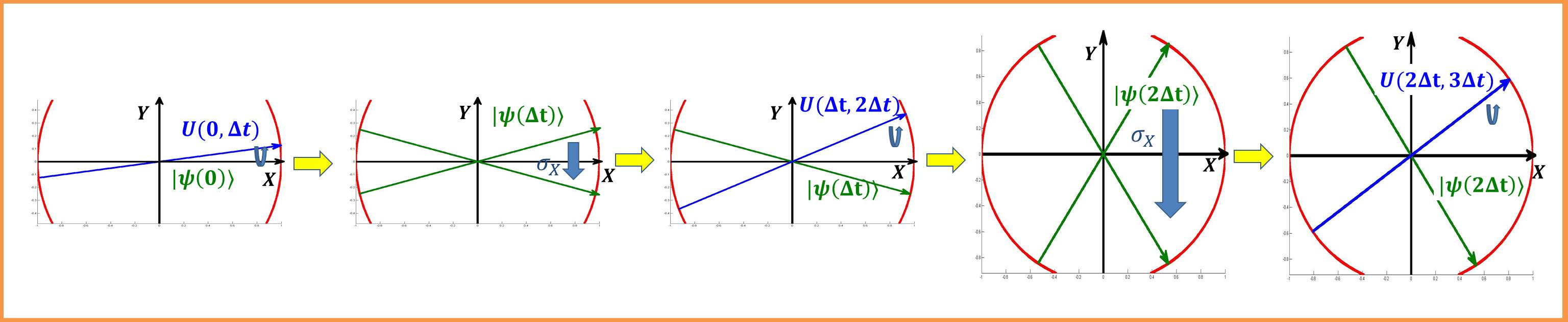}}
\end{center}
\caption{Illustration of the second control method (see section \ref{second}) in Bloch sphere ($x-y$ plane). Top: Adiabatic evolution of the spin under the Hamiltonian of eq. \ref{interaction_Hamiltonian}. The unitary is rotated with a frequency of $2 \delta$ and the state evolves along it. The state undergoes a $\pi$-pulse with a different unitary every $\Delta t$, as $\Delta t=\frac{\pi}{2 \Omega}.$ However these different angles do not accumulate and it rotates with the same frequency as the unitary. Bottom: Now in addition to the unitary evolution a $\pi$-pulse around $\sigma_{X}$ axis is applied every $\Delta t.$ Due  to these $\pi$-pulses the different angles are accumulated, and the rotation angle of the state goes as $\delta \frac{t^{2}}{\Delta t}.$}
\label{phase_acceleration}
\end{figure*}

\subsection {Second control method}
\label{second}
Let us first examine why a standard scaling is achieved in the absence of control. 
It can be seen that the dynamics in the slot $(0,t),$ in the absence of any control, is given by the unitary ( see appendix \ref{appendix:derivation}):
\begin{equation}
U(0,t)=\exp\left(-i\delta \sigma_{Z} t \right)\exp\left(-i\left( -\delta\sigma_{Z}+\Omega \sigma_{X} \right) t \right).
\label{time_ev_1}
\end{equation}
This time evolution clearly cannot lead to an improved FI as it represents the concatenation of two standard rotations. Observe that the unitary for the time slot $(  t,t+\Delta t ) $ reads (see appendix \ref{appendix:derivation}):
\begin{eqnarray}
\begin{split}
&U\left(t,t+\Delta t\right)=\exp\left(-i\delta\sigma_{Z}\Delta t\right) \cdot
\\
&\exp\left(-i\left(-\delta\sigma_{Z}+\Omega\cos\left(2\delta t\right)\sigma_{X}+\Omega\sin\left(2\delta t\right)\sigma_{Y}\right)\Delta t\right).
\label{time_ev_2}
\end{split}
\end{eqnarray} 
In leading order of $\frac{\delta}{\Omega},$ where $\delta \ll \Omega,$  the unitary reads:
\begin{eqnarray}
\begin{split}
& U\left(t,t+\Delta t\right)=\exp\left(-i\delta\sigma_{Z}\Delta t\right) \cdot  \\
& \exp\left(-i\Omega\left(\cos\left(2\delta t\right)\sigma_{X}+\sin\left(2\delta t\right)\sigma_{Y}\right)\Delta t\right).
\end{split}
\end{eqnarray}
Taking $\Delta t=\frac{\pi}{2 \Omega}$ we get:
\begin{equation}
U\left(t,t+\Delta t\right)=i\left(\sigma_{X}\cos\left(2\delta t+\delta\Delta t\right)+\sigma_{Y}\sin\left(2\delta t+\delta\Delta t\right)\right).
\end{equation}

The intuition is clear, in the limit of small $\delta \Delta t$ we expect $U\left(t,t+\Delta t\right)\approx\exp\left(-i\Omega\left(\cos\left(2\delta t\right)\sigma_{X}+\sin\left(2\delta t\right)\sigma_{Y}\right)\Delta t\right).$
Hence $U\left(t,t+\Delta t\right)$ is a rotation around $\sigma_{X}\cos\left(2\delta t+\delta\Delta t\right)+\sigma_{Y}\sin\left(2\delta t+\delta\Delta t\right)$ axis with an angle of $\pi.$ The dynamics therefore can be understood as follows: in each time slot of $\left(t,t+\Delta t\right)$ our state is rotated around $\sigma_{X}\cos\left(2\delta t+\delta\Delta t\right)+\sigma_{Y}\sin\left(2\delta t+\delta\Delta t\right)$ axis with an angle of $\pi.$ The axis of rotation thus rotates, unlike the static Hamiltonian case. It should be noted that this property is not manifested in the state evolution. For example taking $|\downarrow_{x} \rangle$ as an initial state, it will undergo the following evolution:
\begin{eqnarray}
\begin{split}
& U\left(t,0\right)|\downarrow_{x}\rangle=U\left( \left(N-1\right) \Delta t, N\Delta t\right)...U\left(\Delta t,2 \Delta t\right)U\left(0, \Delta t\right)|\downarrow_{x}\rangle \\
& =\cos\left(\delta N\Delta t\right)|\downarrow_{x}\rangle+i\sin\left(\delta N\Delta t\right)|\uparrow_{x}\rangle
\end{split}
\end{eqnarray}
 This adiabatic evolution is illustrated in fig. \ref{phase_acceleration}, showing that the state rotates with the unitaries and therefore the accumulated phase is linear with time. In fact, in each time interval the previous angle of rotation is subtracted from the new angle of rotation, and thus the phase accumulation is not optimal. 
 
 Clearly, if the angles of rotations could be summed up instead of subtracted from each other, then the total acquired phase would be larger, resulting in a better FI. This indeed can be accomplished just by applying a $\pi$-pulse around the $\sigma_{X}$-axis (or $\sigma_{Y}$-axis) every $\Delta t,$ namely by reflecting across the $\sigma_{X}$-axis. The intuition behind this control is clarified in fig. \ref{phase_acceleration}, the angles of rotations are now summed up leading to an accelerated phase accumulation. Taking $|\downarrow_{x}\rangle$ as an initial state we have:
 \begin{eqnarray}
 \begin{split}
 &  U\left( \left(N-1\right) \Delta t, N \Delta t\right)\Pi_{x}...\Pi_{x}U\left(\Delta t, 2 \Delta t\right)\Pi_{x}U\left(0 , \Delta t\right)|\downarrow_{x}\rangle \\
 & =\cos\left(\delta\underset{k=1}{\overset{N}{\sum}} \left( 2k-1 \right) \Delta t\right)|\downarrow_{x}\rangle+i\sin\left(\delta\underset{k=1}{\overset{N}{\sum}}\left( 2k-1 \right) \Delta t\right)|\uparrow_{x}\rangle \\
& =\cos\left(\delta \frac{t^{2}}{\Delta t}  \right)|\downarrow_{x}\rangle+i\sin\left(\delta \frac{t^{2}}{\Delta t} \right)|\uparrow_{x}\rangle,
 \end{split}
 \end{eqnarray}                
where $t$ is the total time. The resulting FI reads: $I=4\left(  \frac{t^{2}}{\Delta t}      \right)^{2}=4 \frac{t^{4}}{\left(\Delta t\right)^{2}}=4 \Omega^{2} t^{4}  \left( \frac{2}{\pi}  \right)^{2},$  hence the accelerated phase accumulation gives rise to a $t^{4}$ scaling. Note that although optimal scaling is achieved, the FI of this method is smaller than the optimal value by a prefactor of $\left(\frac{2}{\pi}\right)^{2}.$ More generally, for $\Delta t=\left(2k+1\right)\frac{\pi}{2\Omega}$ the FI reads: 
\begin{equation} 
I=4\Omega^{2}t^{4}\left(\frac{2}{\pi\left(2k+1\right)}\right)^{2}.
\end{equation}
The scheme works as long as our approximation of $\sqrt{\Omega^{2}+\delta^{2}}t\approx\Omega t,$ holds, which means that the duration of this method is approximately $\frac{\Omega}{\delta^{2}}.$ This presents a significant improvement over the lifetime of the other methods ($\frac{1}{\delta}$). A comparison between the different methods is presented in figure \ref{comparison_methods}(b).
Another advantage, which might be experimentally important, is that the SNR (signal to noise ratio) in this method is much improved over that in other schemes.
Recall that in the first method the effective Hamiltonian scaled as $\Omega \delta t,$ with a limitation of $\delta t \ll 1,$ while here the effective Hamiltonian scales as $\Omega.$

\begin{figure}[b]
\begin{center}
\includegraphics[width=0.4\textwidth]{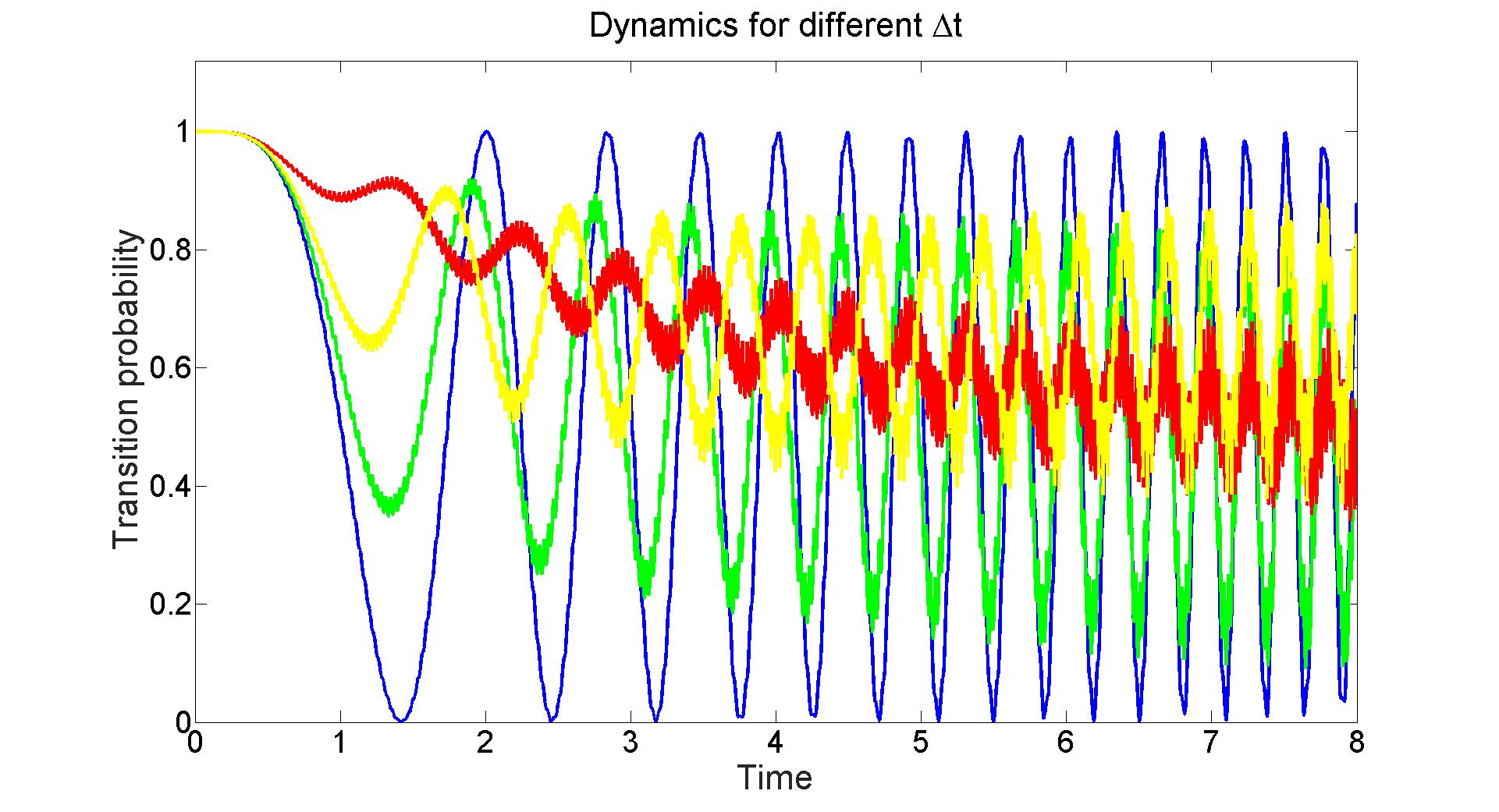}
\includegraphics[width=0.4\textwidth]{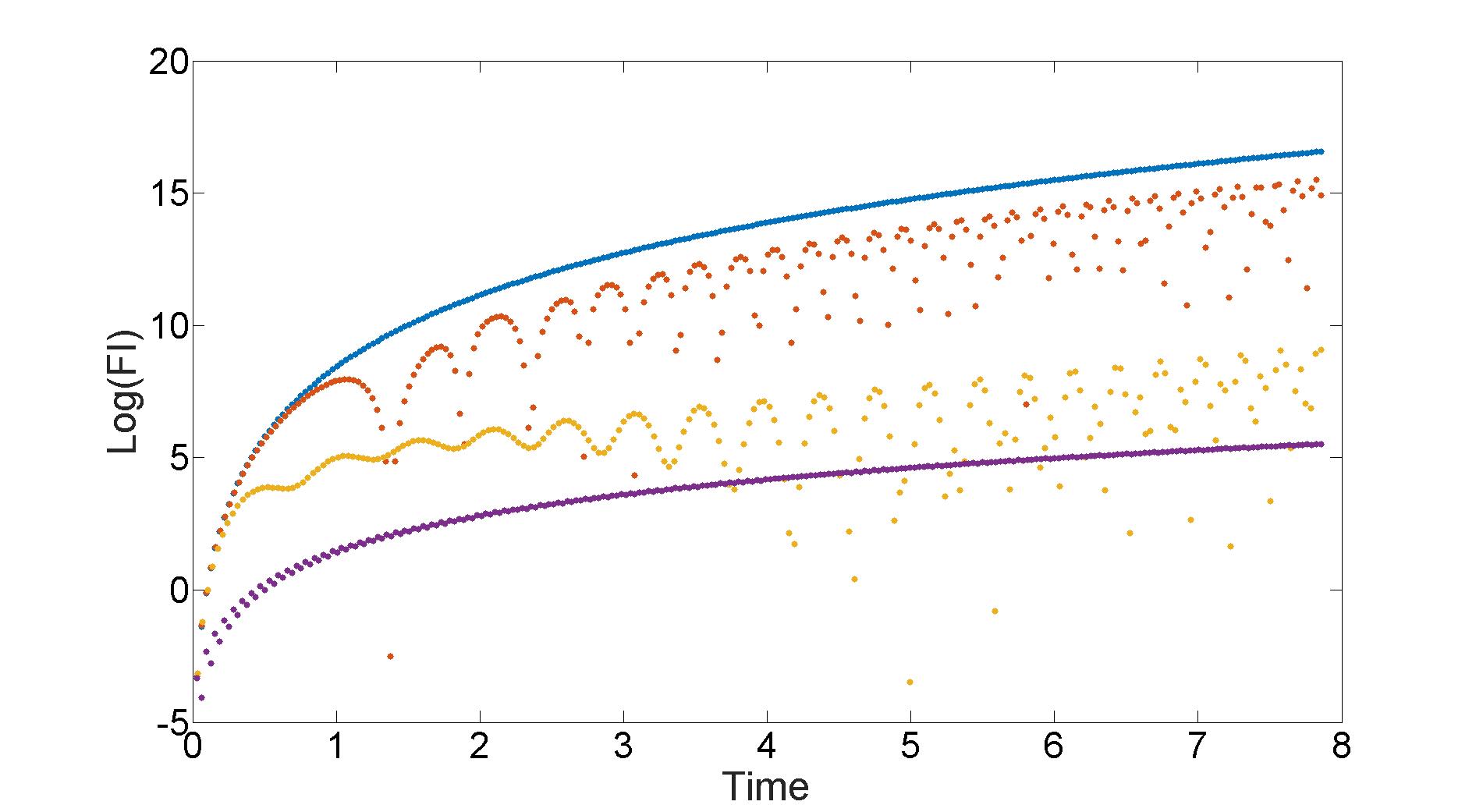}
\end{center}
\caption{Numerical analysis of the FI behavior under different timings of $\pi$-pulses in the second control method (see section \ref{second}). Top: dynamics of transition probability for different timings, where $\Delta t=\frac{\pi}{2\Omega}\text{ (blue)},\frac{\pi}{1.96\Omega}\text{ (green)},\frac{\pi}{1.9\Omega}\text{ (red)},\frac{\pi}{2.06\Omega}\text{ (yellow)}.$ It can be seen that for $\Delta t=\frac{\pi}{2 \Omega}$ perfect $T^{4}$ oscillations are achieved, while for other values the amplitude drops significantly. Bottom: comparison of the FI for different timings: $\Delta t=\frac{\pi}{2\Omega}\text{ (blue)},\frac{\pi}{1.96\Omega}\text{ (orange)},\frac{\pi}{1.8\Omega}\text{ (yellow)}$ and no control (purple).}
\label{numerical_analysis}
\end{figure}

The main limitation of this scheme is the need to know $\Omega$ to a high degree, so that $\pi$-pulses can be applied every $\frac{\pi}{2 \Omega}.$ In order to study the sensitivity to uncertainty in $\Omega$, we shall examine the behavior of the FI for different timings ($\Delta t$) of the $\pi$-pulses. Clearly not every $\Delta t$ leads to a $T^{4}$ scaling: Taking $\Delta t=N \frac {\pi}{\Omega},$ for an integer $N,$ we obtain $U\left(t,t+\Delta t\right)=\exp\left(i\delta\sigma_{Z}\Delta t\right)$ and thus phase acceleration cannot be obtained. For a general $\Delta t$ the analysis becomes more tricky, some numerical results are shown in figure \ref{numerical_analysis}. For $\Delta t=\frac{\pi}{2\Omega}$ the probability equals to $\cos^2\left(\delta\frac{t^{2}}{\tau}\right)$, and thus $T^{4}$ oscillations have a unit amplitude (up to small deviations that go as $\frac{\delta^{2}}{\Omega}t$). These $T^{4}$ oscillations do not disappear for different $\Delta t,$ but their amplitude drops dramatically and vanishes for values of $\frac{N \pi}{\Omega}.$

We note that this analysis is also relevant for $H_{2}$ (eq. \ref{Qdyne_Hamiltonian}). This can be seen by moving to the interaction picture with respect to $\omega' \sigma_{Y}$ and neglecting the fast rotating terms, which yields: $H_{I}=\frac{\Omega}{2}\left(\sigma_{Z}\cos\left(2 \delta t\right)+\sigma_{X}\sin\left(2 \delta t\right)\right).$ In this case the FI drops by a factor of $\frac{1}{4}$ but the lifetime is longer. Note that that if $H_{2}$ is realized by applying an electromagnetic field (as described in ref. \cite{Qdyne}), then we can obtain the desired Hamiltonian by simply changing the polarization of the field to a circular one. This can be done via a configuration which is shown in \cite{mutlu,liu,song,Zhancheng,harvey1963microwave}

\subsection {Unknown initial phase}
 Until now the initial phase of the Hamiltonian was assumed to be known (and taken to be zero), however in most realistic scenarios this is not the case. This is because it is impossible to lock the control to the phase of the signal, which is usually created by an external source.
We now show that an unknown phase does not change dramatically the precision in estimating $\omega$. It can be easily seen that the second method yields an accelerated phase in this case as well.
Given an unknown initial phase $\varphi,$ observe that for $\Delta t=\frac{\pi}{2 \Omega}$ we have $U\left(k\Delta t,\left( k+1 \right) \Delta t\right)=i\left(\sigma_{x}\cos\left(\left(2k+1\right)\Delta t+\varphi\right)+\sigma_{y}\sin\left(\left(2k+1\right)\Delta t+\varphi\right)\right).$ Therefore the transition probability now reads $\sin\left(\delta\left(\frac{t^{2}}{\Delta t}\right)+\frac{t}{\Delta t}\phi\right)^{2}.$ There is still a quadratic phase accumulation, the only difference now is the new term $\frac{t}{\Delta t} \varphi,$ namely the unknown phase. 
It is now easy to see that the FI of $\delta$ is the same: $4\Omega^{2}\left(\frac{2}{\pi}\right)^{2}T^{4}.$ However this quantity is meaningless, as the entire FI matrix should be calculated. Since we consider two unknown parameters, $\delta,\, \varphi ,$ this should be a $2\times2$ matrix, which we denote as $I_{m}.$
Recall that the variance in estimating $\delta$ is now bounded by $\left(I_{m}^{-1}\right)_{\delta,\delta},$ hence the quantity of interest is: $1/\left(I_{m}^{-1}\right)_{\delta,\delta}.$ The FI matrix of $\left(  \delta,\, \varphi  \right)$ now reads:    

\begin{equation}
I=\left(\begin{array}{cc}
I_{\delta,\delta} & I_{\delta,\varphi}\\
I_{\varphi,\delta} & I_{\varphi,\varphi}
\end{array}\right)=4\Omega^{2}\left(\frac{2}{\pi}\right)^{2}\left(\begin{array}{cc}
T^{4} & T^{3}\\
T^{3} & T^{2}
\end{array}\right),
\end{equation} 
 it is easy to see that $I_{m}$ is not invertible, which implies infinite variance. This makes sense as no information about $\delta$ can be obtained due to the unknown phase. In order to retrieve the $T^{4}$ scaling we must have identical probes that feel the same signal and they should have different measurement times or use the same probe twice in time when the phase is constant. For two systems, it can be easily seen that the optimal measurement times are: $T,0.45T ,$ which yields FI of $\sim\frac{\Omega^{2}}{10}\left(\frac{2}{\pi}\right)^{2}T^{4}.$ Thus, $T^{4}$ scaling can be achieved, but with a prefactor of $\frac{1}{40}$ due to the unknown phase.  Determining the optimal measurement times and finding tight bounds of the FI for a large number of systems is left as an open challenge.

As for the first method: the phase acceleration depends on the phase, but only phases of $\phi=\pm\frac{\pi}{2}$ ruin it completely.
It can be seen explicitly: in the first dynamical decoupling method we had: $H_{Ie}=\Omega \sigma_{Y} \sin \left( 2 \delta t+\phi  \right),$ so for $\delta t \ll 1$ it is just: $H_{Ie}=\Omega\left(2\delta t\cos\left(\phi\right)+\sin\left(\phi\right)\right)\sigma_{Y}.$ Hence the phase acceleration term is multiplied by a factor of $\cos (\phi).$ A similar analysis of the FI matrix yield similar results: Optimizing over two measurement times the FI reads $\sim\frac{\Omega^{2}}{10}T^{4} \cos(\phi)^{2}.$
So as expected there is a factor of $\cos(\phi)^{2},$ and the optimization problem for the general case is the same as in the second method.

\section {Finite coherence time and $T^{3}$ scaling} 
 Our analysis did not take into account the effect of noise, namely a finite coherence time of the probe and of the signal. Let $T_{\phi} \, (\tau)$ denote the coherence time of the signal (probe). For a variety of probes, including NV centers, $\tau$ is basically $T_{2}.$ This scaling is obviously important in the case of $T_{\phi}\leq\tau.$\\ 
 
 It becomes less obvious in the other regime, in which the coherence time of the probe is much shorter than the coherence time of the signal, namely $\tau \ll T_\phi \approx T,$ where $T$ is total time. The main observation regarding this regime is that either with or without coherent control, the FI scales as $T^{3},$ which is still an improvement compared to the usual scaling of $T^{2}.$ 
 
 Let us show this first for the Hamiltonian of the first kind. 
We can get an intuition to this scaling by examining the case of $\omega \ll \Omega,$ where we have seen that for $\tau=\frac{\pi}{2\left(2N+1\right)}$ the unitary in the slot $\left(t,t+\tau\right)$ is given by: $U=i\left(\sigma_{X}\cos\left(\delta\left(2t+\tau \right)\right)+\sigma_{Y}\sin\left(\delta\left(2t+\tau \right)\right)\right).$ 
Therefore setting the probe to an eigenstate of $\sigma_{X}$ and performing a measurement in this slot yields a transition probability of $\sin^{2}\left(\delta\left(2t+\tau\right)\right),$
which accounts for FI of $4\left(2t+\tau \right)^{2}$. Performing $\frac{T}{\tau}$ measurements (starting at $t=0$) yields the following FI (for $\tau=\frac{\pi}{2\left(2N+1\right)}$): $4\underset{t}{\sum}\left(2t+\tau\right)^{2}\approx16\frac{T^{3}}{\tau}.$ Therefore this scaling comes from the fact that each single measurement at $t$ yields FI that scales as $t^{2},$ so that the total FI scales as $T^{3}.$ In the following we perform a more precise and general analysis:  
The best FI achievable for $\frac{T}{\tau}$ measurements would be the sum of the QFIs of these measurements. Recall (eq. \ref{QFI}) that the QFI of a measurement in the slot $\left(t,t+\tau\right)$ reads: $\text{Var}\left(iU_{t}^{\dagger}\overset{\cdot}{U_{t}}\right),$ where $U_{t}$
is the time evolution unitary in this interval and the derivative is with respect to $\omega$. Therefore the FI reads: 
\begin{equation}
I_{\text{tot}}=\underset{t}{4\sum}\text{Var}\left(iU_{t}^{\dagger}\overset{\cdot}{U_{t}}\right),
 \end{equation}
where $U_{t}$ is exactly the same as in eq. \ref{time_ev_2} (when $\delta$ is replaced by $\omega$). For $\tau \ll t$ we get that (see appendix \ref{appendix:variance}):
\begin{equation}
\text{Max}\left[\text{Var}\left(iU_{t}^{\dagger}\overset{\cdot}{U_{t}}\right)\right]\approx\frac{4\Omega^{2}t^{2}}{\Omega^{2}+\omega^{2}}\sin\left(\sqrt{\omega^{2}+\Omega^{2}}\tau \right)^{2}.
\end{equation}
Therefore the maximal FI in that case reads:
\begin{eqnarray}
\begin{split}
&I_{\text{tot}}\approx\underset{t}{\sum}\frac{16\Omega^{2}t^{2}}{\omega^{2}+\Omega^{2}}\sin\left(\sqrt{\omega^{2}+\Omega^{2}}\Delta t\right)^{2} \approx \\
&\frac{16\Omega^{2}}{\omega^{2}+\Omega^{2}}\sin\left(\sqrt{\omega^{2}+\Omega^{2}}\tau \right)^{2}\frac{T^{3}}{3\tau}.
\label{FI_first_kind}
\end{split}
\end{eqnarray}     
Obviously too short $\tau$ leads to a poor FI, while for too long $\tau$ oscillations lead to a loss of information and thus a suboptimal FI. Mathematically this is just the tradeoff between $\sin\left(\sqrt{\omega^{2}+\Omega^{2}}\tau\right)^{2}$ and $\frac{1}{\tau}$ in eq. \ref{FI_first_kind}, which suggests that the FI gets an optimum for a certain $\tau.$ This means that in the absence of control it is not always preferable to use the entire coherence time, but rather a shorter measurement period.   
This optimum $\tau$ is found to be $\tau=\frac{1.16}{\sqrt{\omega^{2}+\Omega^{2}}},$ thus the optimal FI reads: $3.86\frac{\Omega^{2}}{\sqrt{\omega^{2}+\Omega^{2}}}T^{3}.$ This already suggests that for coherence time longer than $\frac{1.16}{\sqrt{\omega^{2}+\Omega^{2}}},$ coherent control may be useful. \\  	      

To see this, Recall that with coherent control an effective Hamiltonian: $H_{e}=2\Omega \delta t \sigma_{Y}$ can be obtained. In this case the FI for the interval $\left( t,t+\tau \right)$ is $4\Omega^{2}\left(\left(t+\tau\right)^{2}-t^{2}\right)^{2},$ and the total FI reads:
 \begin{equation} 
I_{tot}=\underset{t}{\sum}4\Omega^{2}\left(\left(t+\tau\right)^{2}-t^{2}\right)^{2}\approx\frac{16}{3}\left(\Omega^{2}\tau\right)T^{3},
\end{equation}
where, again, $\tau \leq T_{2}.$ 
This again yields a $T^{3}$ scaling, but for long enough $\tau$ it leads to an improvement that goes as $\sqrt{\Omega^{2}+\omega^{2}}\tau.$

\begin{figure}[h]
\begin{center}
\includegraphics[width=0.36\textwidth]{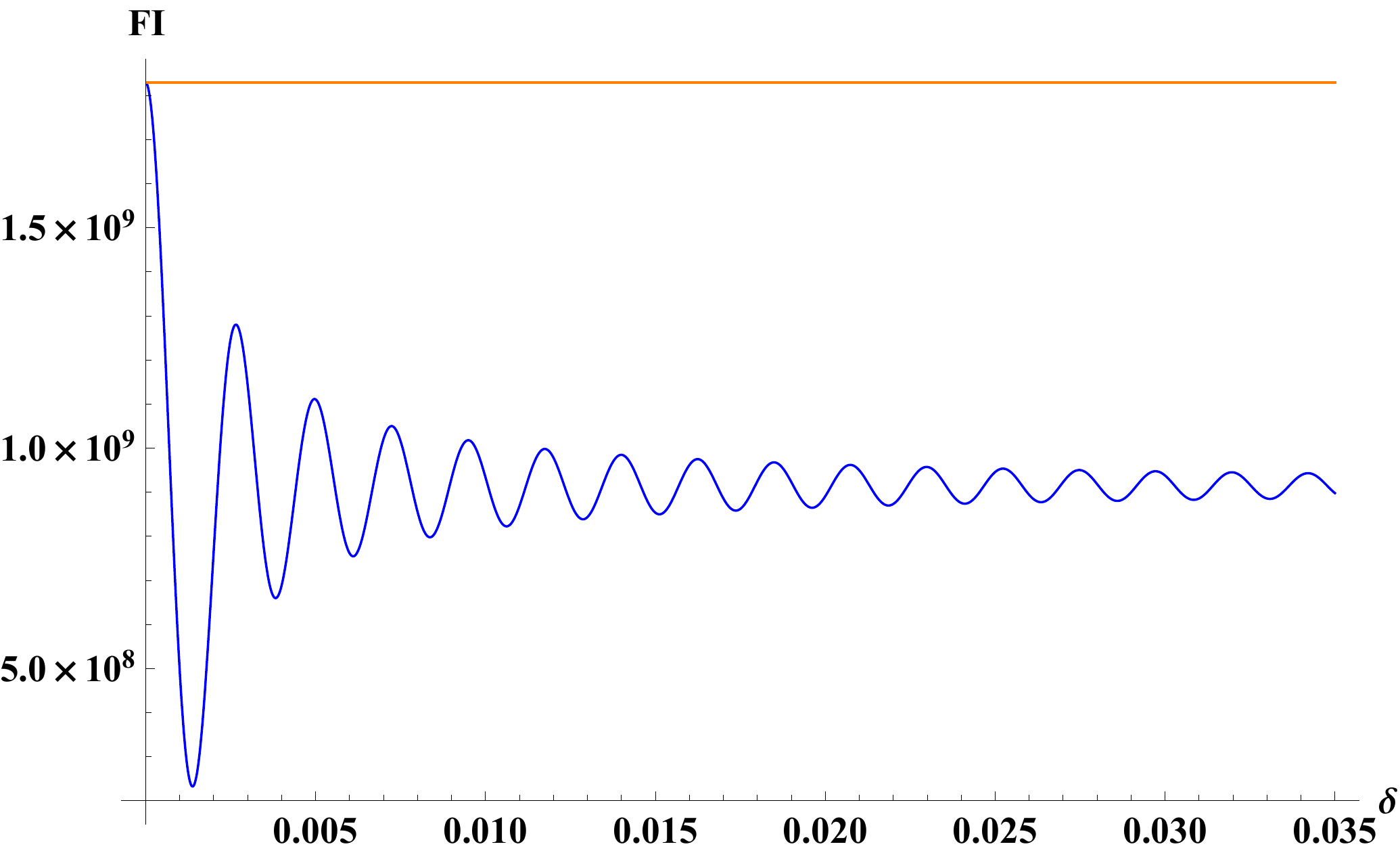}
\includegraphics[width=0.35\textwidth]{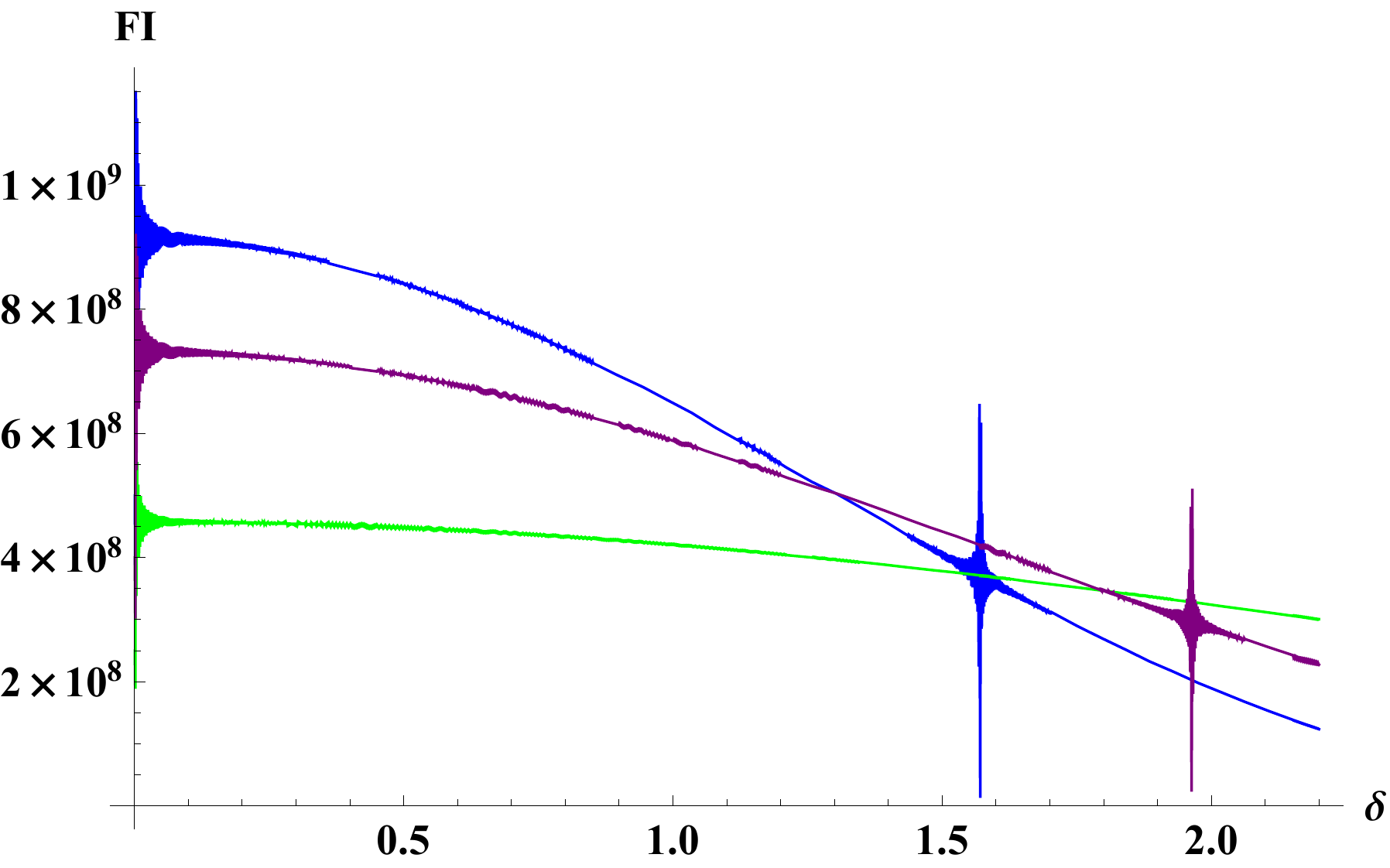}
\end{center}
\caption{Analysis of the FI as a function of time for a given measurement period $\tau$. Top: Behavior of the FI in the regime of $\delta \tau \ll 1.$ Observe that the FI coincides with the ultimate limit (orange curve) for $\delta T \ll 1$ and then oscillates and drops to half of this limit. Bottom: The FI for larger $\delta$ and different values of $\tau:\,1 \, \text{(blue)}, \, 0.8 \, \text{(purple)}, \, 0.5 \, \text{(green)} \, [\frac{1}{\Omega}].$ The FI goes as $I_{\text{tot}}=8\left(\frac{\Omega}{\delta}\right)^{2}\frac{T^{3}}{3\tau}\sin \left( \delta  \tau \right),$ this serves as a good approximation except for $\delta=0 \, , \frac{\pi}{2 \delta}$ in which there are sharp oscillations. $\tau$ is the optimal measurement period only for $\delta<\frac{1.16}{\tau}.$ }
\label{last}
\end{figure}

The picture is a bit different for the Hamiltonian $H=\Omega \sigma_{Z} \sin \left( 2 \omega t  \right). $ We know that by applying the appropriate control the frequency $\omega$ is changed to $\delta$ and for $\delta T \ll 1,$ a $T^{4}$ scaling is achieved.
Therefore, let us focus on $H=\Omega \sigma_{Z} \sin \left( 2 \delta t  \right) ,$ and examine how the FI changes change as a function of $\delta$ for a given single experiement time of $\tau.$ 
The total FI reads: 
\begin{eqnarray}
\begin{split}
&I_{\text{tot}}=\frac{\Omega^{2}}{\delta^{4}}\cdot\underset{t}{\sum}[2\delta t\sin\left(2\delta t\right)-2\delta\left(t+\tau\right)\sin\left(2\delta\left(t+\tau\right)\right)\\
&+\cos\left(2\delta t\right)-\cos\left(2\delta\left(t+\tau\right)\right)]^{2}
\end{split}
\end{eqnarray}
It can be seen that for $\delta \tau \ll 1$ the FI is given by:
\begin{equation}
I_{\text{tot}}=16\Omega^{2}\tau\left(\frac{T^{3}}{6}+\frac{T\cos\left(4\delta T\right)}{16\delta^{2}}+\frac{\left(8\delta^{2}T^{2}-1\right)\sin\left(4\delta T\right)}{64\delta^{3}}\right).
\end{equation} 
Hence for $\delta T \ll 1$ we get $\frac{16}{3}\Omega^{2}\tau T^{3},$ which is limited by  $\frac{16}{3}\Omega^{2} T_2 T^{3}.$
In the regime where $\delta T$ is no longer small but still $\delta \tau \ll 1,$ the FI drops by a factor of $2$ to $\frac{16}{6}\Omega^{2}\tau T^{3}.$ This is just due to the factor of $\cos^{2} \left( \phi \right)$ that is added in an arbitrary phase $\phi.$ Now for larger $\delta,$ where $\delta \tau$ is no longer small, the FI to a good approximation reads:
\begin{equation}
I_{\text{tot}}=8\left(\frac{\Omega}{\delta}\right)^{2}\frac{T^{3}}{3\tau} \sin\left(\delta\tau\right)^{2}.
\label{Qdyne_big_delta}
\end{equation}
 Note that this expression is very similar to the one in eq. \ref{FI_first_kind}, and similarly if $\tau$ is too large it is no longer optimal to perform a measurement every $\tau,$ and shorter measurement periods are preferable. Unlike with the Hamiltonian of the first kind, the optimal measurement period depends only on $\delta.$ It can be seen that $\tau$ is optimal as long as $\delta<\frac{1.16}{\tau},$ and for larger $\delta$ the optimal measurement period is $\frac{1.165}{\delta}$, as shown in fig.\ref{last}. Plugging this into eq. \ref{Qdyne_big_delta}, we get: 
 \begin{equation}
 I_{\text{tot}}\approx1.93\frac{\Omega^{2}}{\delta}T^{3},
  \end{equation}   
therefore in this regime, if the measurement period is chosen wisely the FI drops as $\frac {1}{\delta}.$

{\it Conclusions and outlook---}   This paper describes novel schemes designed to achieve $T^{4}$ scaling of the FI. A detailed analysis of the schemes verifies that these scaling persist for extended times. As the proposed $T^4$ methods are adaptive the extension of time is crucial for experimental realization. We anticipate that the use of these schemes will find applications in nano imaging and in atomic clocks protocols. It is still an open question as to whether $T^4$ scaling can be reached in a non adaptive way.
It is noteworthy that the resolution of the $T^4$ method scales with $(\Omega T_2)^2,$ meaning that it improves quadratically with the strength of the signal.

\appendix
\section{Deravition of the time evolution}
\label{appendix:derivation}
The goal of this section is to derive the unitaries in eqs. \ref{time_ev_1},\ref{time_ev_2}. Let us begin with the unitary in eq. \ref{time_ev_1}, which is the time evolution obtained from $H=\Omega \left( \sigma_{X}\cos\left(2\delta t\right)+\sigma_{Y}\sin\left(2\delta t\right) \right)$ in the interval $\left(0,t\right).$ Observe that $H$ can be obtained from $H_{s}=-\delta \sigma_{Z}+\Omega \sigma_{X},$ by moving to the interaction picture with respect to $H_{0}=-\delta\sigma_{Z},$ as $e^{-i\delta\sigma_{Z}t}\sigma_{X}e^{i\delta\sigma_{Z}t}=\sigma_{X}\cos\left(2\delta t\right)+\sigma_{Y}\sin\left(2\delta t\right).$ 
This immediately implies, from the definition of interaction picture, that: 
\begin{eqnarray}
\begin{split}
&U\left(0,t\right)=\exp\left(iH_{0}t\right)U_{s}\left(0,t\right) = \\
&\exp\left(-i\delta\sigma_{Z}t\right)\exp\left(-i\left(-\delta\sigma_{Z}+\Omega\sigma_{X}\right)t\right),
\end{split}
\end{eqnarray}     
which justifies eq. \ref{time_ev_1}.

It is now straightforward to get $U\left(t,t+\Delta t\right),$ since solving the differential equation for $H=\Omega\left(\sigma_{X}\cos\left(2\delta t\right)+\sigma_{Y}\sin\left(2\delta t\right)\right),$ in the slot $\left(t,t+\Delta t\right)$ is the same as solving for $H'=\Omega\left(\sigma_{X}\cos\left(\phi+2\delta t\right)+\sigma_{Y}\sin\left(\phi+2\delta t\right)\right)$ in the slot $\left(0,\Delta t\right),$ where $\phi=2\delta t.$ This means that $U\left(t,t+\Delta t\right)=U'\left(0,\Delta t\right),$ where $U'$ is the unitary that corresponds to $H'.$ Using the same technique as above, we can readily find:
\begin{eqnarray}
\begin{split}
&U'\left(0,\Delta t\right)=\exp\left(-i\delta\sigma_{Z}\Delta t\right)\\
&\exp\left(-i\left(-\delta\sigma_{Z}+\Omega\left(\sigma_{X}\cos\left(\phi\right)+\sigma_{Y}\sin\left(\phi\right)\right)\right)\Delta t\right).
\end{split}
\end{eqnarray}  
Inserting $\phi=2\delta t,$ we can conclude that:
\begin{eqnarray}
\begin{split}
&U\left(t,t+\Delta t\right)=\exp\left(-i\delta\sigma_{Z}\Delta t\right)\\
&\exp\left(-i\left(-\delta\sigma_{Z}+\Omega\left(\sigma_{X}\cos\left(2\delta t\right)+\sigma_{Y}\sin\left(2\delta t\right)\right)\right)\Delta t\right).
\end{split}
\end{eqnarray} 

\section{FI without control}
\label{appendix:variance} 
Considering $H=\Omega\left(\sigma_{X}\cos\left(2\omega t\right)+\sigma_{Y}\sin\left(2\omega t\right)\right)$ and a coherence time of $\tau,$ the maximal FI in the absence of 
coherent control is the sum of the maximal QFI's of $\frac{T}{\tau}$ measurements. Which means: 
\begin{equation}
\text{FI}=4\underset{t}{\sum}\text{Max}\left[\text{Var}\left(iU_{t}^{\dagger}\overset{\cdot}{U}_{t}\right)\right],
\end{equation}
where $U_{t}$ is the unitary in the slot $\left(t,t+\tau\right)$ and we have used the fact that the QFI of a measurement in this slot is $\text{Var}\left(iU_{t}^{\dagger}\overset{\cdot}{U}_{t}\right).$
To accomplish this calculation we need to find $\text{Max}\left[\text{Var}\left(iU_{t}^{\dagger}\overset{\cdot}{U_{t}}\right)\right]$ to every $t,$ where $U_{t}$ is given by eq. \ref{time_ev_2}. 
Since we are dealing with the limit of $\tau \ll T,$ and are interested in the $T^{3}$ scaling we shall keep only terms that go as $t^{2}$ in $\text{Max}\left[\text{Var}\left(iU_{t}^{\dagger}\overset{\cdot}{U_{t}}\right)\right],$ therefore we keep only the terms that go as $t$ in $iU_{t}^{\dagger}\overset{\cdot}{U}_{t}.$
These terms are: 
\begin{eqnarray}
\begin{split}
&\exp\left(-i\omega\sigma_{z}\tau+i\Omega\left(\sigma_{x}\cos\left(2\omega t\right)+\sigma_{y}\sin\left(2\omega t\right)\right)\tau\right)\cdot \\
&\left(\sin\left(\sqrt{\omega^{2}+\Omega^{2}}\tau\right)\frac{2\Omega t}{\sqrt{\omega^{2}+\Omega^{2}}}\left(-\sigma_{X}\sin\left(2\omega t\right)+\sigma_{Y}\cos\left(2\omega t\right)\right)\right).
\label{operator}
\end{split}
\end{eqnarray}  
Recall that
\begin{equation}
 \text{Max}\left[\text{Var}\left(iU_{t}^{\dagger}\overset{\cdot}{U}_{t}\right)\right]=\frac{\left(\lambda_{\text{max}}-\lambda_{\text{min}}\right)^{2}}{4},
 \label{var}
\end{equation}
   where $\lambda_{\text{max}} \left( \lambda_{\text{min}} \right)$ is the maximal (minimal) eigenvalue of $iU_{t}\overset{\cdot}{U}_{t}^{\dagger}.$ So we need to find the eigenvalues of the operator in eq. \ref{operator}.  
 To this end we shall denote $-\omega\sigma_{Z}+\Omega\left(\sigma_{X}\cos\left(2\omega t\right)+\sigma_{Y}\sin\left(2\omega t\right)\right)$ as $A\sigma_{\varphi}$ and $-\sigma_{X}\sin\left(2\omega t\right)+\sigma_{Y}\cos\left(2\omega t\right)$ as $\sigma_{\theta}.$
 In this notation the operator in eq. \ref{operator} reads $\frac{2\Omega t\sin\left(\sqrt{\omega^{2}+\Omega^{2}}\tau\right)}{\sqrt{\Omega^{2}+\omega^{2}}}\exp\left(iA\sigma_{\varphi}\tau\right)\sigma_{\theta}.$ Observe now that since $\sigma_{\theta}$ and $\sigma_{\varphi}$ are in orthogonal directions we have:
\begin{equation}
\exp\left(iA\sigma_{\varphi}\tau\right)\sigma_{\theta}=\exp\left(iA\sigma_{\theta}\tau/2\right)\sigma_{\theta}\exp\left(-iA\sigma_{\theta}\tau/2\right).
\end{equation}
This implies that $\frac{2\Omega t\sin\left(\sqrt{\omega^{2}+\Omega^{2}}\tau\right)}{\sqrt{\Omega^{2}+\omega^{2}}}\exp\left(iA\sigma_{\varphi}\tau\right)\sigma_{\theta}$ has the same eigenvalues as $\frac{2\Omega t\sin\left(\sqrt{\omega^{2}+\Omega^{2}}\tau\right)}{\sqrt{\Omega^{2}+\omega^{2}}}\sigma_{\theta},$ namely its eigenvalues are $\pm\frac{2\Omega t\sin\left(\sqrt{\omega^{2}+\Omega^{2}}\tau\right)}{\sqrt{\Omega^{2}+\omega^{2}}}.$
Therefore we obtain that:
\begin{eqnarray}
\begin{split}
&\text{FI}\approx \underset{t}{\sum}\frac{16\Omega^{2}t^{2}}{\Omega^{2}+\omega^{2}}\sin\left(\sqrt{\omega^{2}+\Omega^{2}}\tau\right)^{2}\\
&\approx \frac{16\Omega^{2}}{\left(\Omega^{2}+\omega^{2}\right)}\sin\left(\sqrt{\omega^{2}+\Omega^{2}}\tau\right)^{2} \frac{T^{3}}{3\tau}.
\end{split}
 \end{eqnarray}

\vspace{2mm}
{\it Acknowledgements ---}  A. R. acknowledges the support of the Israel Science Foundation(grant no. 1500/13), the support of the European commission (STReP EQUAM Grant Agreement
No. 323714), EU Project DIADEMS, the Marie Curie Career Integration Grant (CIG) IonQuanSense(321798), the Niedersachsen-Israeli Research Cooperation Program and DIP program (FO 703/2-1). This project has received funding from the European Union Horizon 2020(Hyperdiamond).

\baselineskip=12pt

\end{document}